\documentclass[letterpaper, 10 pt, conference]{ieeeconf}
\IEEEoverridecommandlockouts                              

\usepackage{amsmath,amsfonts,amssymb}%
\usepackage{bm}%
\usepackage{graphicx}%
\usepackage{cases}%
\usepackage{cite,url,citesort}
\usepackage{verbatim,latexsym}

\newtheorem{definition}{Definition}
\newtheorem{assumption}{Assumption}
\newtheorem{theorem}{Theorem}
\newtheorem{remark}{Remark}

\newtheorem{lemma}{Lemma}
\newtheorem{corollary}{Corollary}

\begin{document}
%
\title{\LARGE \bf
 Coherent Robust $H^{\infty}$ Control of Uncertain Linear Quantum Stochastic Systems}

\author{Chengdi Xiang, Ian R. Petersen and Daoyi Dong
\thanks{Chengdi Xiang, Ian R. Petersen and Daoyi Dong are with the School of Engineering and Information Technology, University of New South Wales at the Australian Defence Force Academy, Canberra ACT 2600, Australia. {\tt\small \{elyssaxiang, i.r.petersen, daoyidong\}@gmail.com}}
}


\maketitle

\begin{abstract}
This paper considers a class of uncertain linear quantum systems subject to uncertain perturbations in the system Hamiltonian. We present a method to design a coherent robust $H^\infty$ controller so that the closed loop system is robustly stable and achieves a prescribed level of disturbance attenuation with all the admissible uncertainties. An illustrative example shows that for the given system, the method presented in this paper has improved performance over the existing quantum $H^\infty$ control results without considering uncertainty.
\end{abstract}

\IEEEpeerreviewmaketitle



\section{Introduction}
Quantum control has been an active research area with wide applications in quantum optics, quantum communication, quantum computation and other quantum technologies \cite{Altafini}, \cite{Wiseman}, \cite{daoyi}. In this area, robustness is counted as one of the most important issues in quantum control systems because quantum systems are unavoidably subject to all kinds of disturbances and uncertainties \cite{matt2008}, \cite{sliding_mode}, \cite{Dong2015}. Feedback control has been recognized as the most effective method to enhance robustness in classical control systems and quantum feedback control theory has also been developed for dealing with disturbances and uncertainties in quantum systems [7]-[21]. For example, to attenuate the influence of disturbance signals, references \cite{matt2008} and \cite{aline2010} focused on finding a controller to bound the influence of the disturbance input signal on the performance output signal using $H^\infty$ synthesis.
Reference \cite{matt2008} considered the quadrature form of the quantum system variables and presented a systematic method to design quantum $H^\infty$ controllers for a class of linear stochastic quantum systems. In order to simplify the work of \cite{matt2008}, paper \cite{aline2010} considered a coherent $H^\infty$ control problem for a class of quantum system called `passive' systems defined in terms of annihilation operators only. To deal with the issue of uncertainty in quantum system models, papers \cite{Chengdi1} and \cite{Chengdi2} designed a Hamiltonian controller to allow for norm bounded uncertainty in the nominal system, to make the closed system robustly stable and achieve a better performance than the system without a controller. Reference \cite{shaiju} also presented a method to synthesize classical guaranteed cost controllers for uncertain quantum systems. Reference \cite{shaiju} addressed not only the issue of robustness but also the issue of LQG performance.

However, little attention in quantum feedback control has been paid to the $H^\infty$ problem with parameter uncertainties in the quantum system model. Hence, in this paper, we consider a class of  linear quantum systems subject to both perturbation uncertainty and input disturbances. The paper aims to design a quantum controller to robustly stabilize the uncertain quantum system as well as to guarantee a prescribed level of disturbance attenuation in the $H^\infty$ sense for the closed loop system with all admissible uncertainties.

In the classical (non-quantum) case, robust $H^\infty$ control for uncertain linear systems has been addressed widely, e.g., \cite{lihua1992}, \cite{kemin} and the relationship between $H^\infty$ optimization and the robust stabilization of uncertain linear systems has been established \cite{kemin}. Hence, in this paper, we adopt some results \cite{lihua1992} on classical control systems to build a relationship between a coherent robust $H^\infty$ control problem and a scaled $H^\infty$ control for a system without parameter uncertainty.

In this paper, we aim to design a coherent dynamic output feedback controller. Here coherent refers to the fact that the controller itself is a quantum system. In this way, a physical realizability condition needs to be considered. The objective of the controller is to reduce the effect of the disturbance input on the controlled output, uniformly in an $H^\infty$ sense.

We begin in Section \ref{section2} by introducing a general class of linear quantum stochastic systems and presenting the physical realizability condition for a given quantum system. In Section \ref{section3}, we introduce unknown perturbations in the system Hamiltonian and parameterize the perturbation in terms of norm bounded uncertainty in the state matrix. In Section \ref{section4}, we introduce some theory to establish relation between a robustly strict bounded real lemma of the uncertain quantum system and a strict bounded real lemma of the scaled $H^\infty$ system. In Section \ref{section5}, we provide a systematic method to design an output feedback quantum controller that satisfies the specified $H^\infty$ requirement. In Section \ref{illustrative example}, we provide an example to demonstrate the theory that has been developed in this paper and compare  the $H^\infty$ performance using the method in this paper with the results in \cite{matt2008}. Conclusions are presented in Section \ref{conclusion}.

\section{Linear Quantum Stochastic Systems}\label{section2}
We consider a class of linear quantum system described by the following non-commutative stochastic differential equations \cite{matt2008}:
\begin{equation}\label{linear}
\begin{split}
dx(t)=&Ax(t)dt+Bdw(t); \quad x(0)=x_0\\
dy(t)=&Cx(t)dt+Ddw(t),
\end{split}
\end{equation}
where $A\in\mathbb{R}^{n\times n}$, $B\in\mathbb{R}^{n\times n_w}$, $C\in\mathbb{R}^{n_y\times n}$ and $D\in\mathbb{R}^{n_y\times n_w}$. Also, $n$, $n_w$, $n_y$ are positive integers and $x(t)=[x_1(t)\quad ...\quad x_n(t)]^T$ is a vector of self-adjoint possible non-commutative system variables.

The initial system variables $x(0)=x_0$ satisfy the commutation relations
\begin{equation}
[x_j(0),x_k(0)]=2i\Theta_{jk}, \quad j,k=1,...,n,
\end{equation}
where $\Theta$ is a real antisymmetric matrix and $\Theta_{jk}$ is the corresponding element of $\Theta$.
Also, the vector $x(t)$ is required to satisfy the canonical commutation relations
\begin{equation}
[x(t),x(t)^T]=x(t)x(t)^T-(x(t)x(t)^T)^T=2i\Theta_{n}
\end{equation}
where  $\Theta_n=\text{diag}_{\frac{n}{2}}(J)=\text{diag}\underbrace{(J,J...,J)}_{\frac{n}{2}}$ for an even number $n$ and $i=\sqrt{-1}$. Here, $J$ denotes the real skew-symmetric $2\times2$ matrix
\begin{equation}
J=\left[\begin{array}{c c} 0&1\\-1&0\end{array}\right].
\end{equation}

The vector quantity $w$ denotes the input signals and is assumed to satisfy the decomposition
\begin{equation}
dw(t)=\beta_w(t)dt+d\tilde{w}(t)
\end{equation}
where $\beta_w(t)$ is a self-adjoint adapted process and $\tilde{w}(t)$ is the noise part of $w(t)$ (see \cite{parthasarathy}). The process $\beta_w(t)$ represents variables of other systems which may be passed to the system (\ref{linear}) via an interaction. The noise $\tilde{w}(t)$ is a vector of quantum Weiner processes with Ito table \cite{parthasarathy}
\begin{equation}
d\tilde{w}(t)d\tilde{w}^T(t)=F_{\tilde{w}}dt
\end{equation}
where $F_{\tilde{w}}=\text{diag}_{\frac{n_w}{2}}(I+iJ)$ is a non-negative definite Hermitian Ito matrix. Hence, the commutation relations for the noise components are shown in the following
\begin{equation}
\begin{split}
[d\tilde{w}(t),d\tilde{w}^T(t)]&=d\tilde{w}(t)d\tilde{w}^T(t)-(d\tilde{w}(t)d\tilde{w}^T(t))^T\\
&=2T_{\tilde{w}}dt
\end{split}
\end{equation}
where we define $S_{\tilde{w}}=1/2(F_{\tilde{w}}+F_{\tilde{w}}^T)$, $T_{\tilde{w}}=1/2(F_{\tilde{w}}-F_{\tilde{w}}^T)$ so that $F_{\tilde{w}}=S_{\tilde{w}}+T_{\tilde{w}}$.
The noise processes can be represented as operators on an appropriate Fock space.

Since $\beta_w(t)$ represents an adapted process, we require $\beta_w(0)$ is an operator on a Hilbert space distinct from that of $x_0$ and the noise processes. We assume that $\beta_w(t)$ commutes with $x(t)$ for all $t\geq0$. Also, we denote $\beta_w(t)$ commutes with $d\tilde{w}(t)$ for all $t\geq0$. Moreover, a property of the Ito increments is that $d\tilde{w}(t)$ commutes with $x(t)$.

The system (\ref{linear}) represents the dynamics of a meaningful physical system if and only if (see \cite{matt2008}):
\begin{equation}\label{preserve}
A\Theta_n+\Theta_nA^T+B\Theta_{n_w}B^T=0
\end{equation}
\begin{equation}\label{preserve2}
BD^T=\Theta_nC^T\Theta_{n_y}
\end{equation}
where equation (\ref{preserve}) preserves the commutation relation; i.e., for $[x_i(0),x_j(0)]=2i\Theta_{ij}$, we always have $[x_i(t),x_j(t)]=2i\Theta_{ij}$ for all $t\geq0$.

\section{Uncertain Linear Quantum System}\label{section3}
In this paper, we consider a class of linear stochastic quantum systems subject to unknown perturbations in the system Hamiltonian.

In order to proceed, we introduce some notation. The symbol $P_m$ describes a $2m\times 2m$ permutation matrix. An $2m\times 2m$ permutation matrix is a full-rank real matrix whose columns consist of standard basis vectors for $\mathbb{R}^{2m}$ such that $P_m^T\left[\begin{array}{c c c c}a_1&a_2&...&a_{2m} \end{array}\right]^T=\left[\begin{array}{c c c c c c c}a_1&a_{m+1}&a_2&a_{m+2}&...&a_m&a_{2m} \end{array}\right]^T$. Also, we have $N_w=(n_w/2)$ and $N_y=(n_y/2)$,
\begin{equation}
M=\frac{1}{2}\left[\begin{array}{c c}1&i\\1&-i \end{array}\right]
\end{equation}
and $\Gamma=P_{N_w}\text{diag}_{N_w}(M)$.

Now we need to introduce an $(S,L,H)$ framework  to define a quantum system \cite{matt2009}.
The Hamiltonian operator $H$ represents self-energy of the system and is in the form of
\begin{equation}
H=\frac{1}{2}x^TRx
\end{equation}
where $R$ is a real symmetric Hamiltonian matrix with dimension $n\times n$.
The coupling operator $L$ is of the form
\begin{equation}
L=\Lambda x
\end{equation}
where $\Lambda$ is a complex-valued coupling matrix with dimension $N_w \times n$.

In this case, the matrices $A, B, C, D$ are given by \cite{matt2008}
\begin{equation}
\begin{split}
A=&2\Theta(R+\Im(\Lambda^\dag \Lambda))\\
B=&2i\Theta[-\Lambda^\dag\quad \Lambda^T]\Gamma\\
C=&P^T_{N_y}\left[\begin{array}{c c} \Sigma_{N_y}&0_{N_y\times N_w}\\0_{N_y\times N_w}&\Sigma_{N_y}\end{array}\right]\left[\begin{array}{c}
\Lambda+\Lambda^\#\\-i\Lambda+i\Lambda^\#\end{array}\right]\\
D=&P^T_{N_y}\left[\begin{array}{c c} \Sigma_{N_y}&0_{N_y\times N_w}\\0_{N_y\times N_w}&\Sigma_{N_y}\end{array}\right]P_{N_w}\\
=&\left[\begin{array}{c c}I_{n_y\times n_y}&0_{n_y\times (n_w-n_y)} \end{array}\right]
\end{split}
\end{equation}
where $\Sigma_{N_y}=\left[\begin{array}{c c}I_{N_y\times N_y}&0_{N_y\times (N_w-N_y)} \end{array}\right]$.

Since the nominal quantum system is subject to uncertain perturbations in the system Hamiltonian, the quadratic perturbation Hamiltonian is assumed in the following form
\begin{equation}
H_{\text{perturbation}}=\frac{1}{2}x^TE^T\Delta Ex,
\end{equation}
where $\Delta\in\mathbb{R}^{m\times m}$ is an uncertain norm bounded real matrix satisfying $\Delta^T=\Delta$ and $\Delta^2\leq I$, and $E\in\mathbb{R}^{m\times n}$.

The uncertain linear quantum system under consideration is described as follows
\begin{equation}\label{original}
\begin{split}
dx(t)=&(A+2\Theta E^T\Delta E)x(t)dt+Bdw(t)\\
dy(t)=&Cx(t)dt+Ddw(t).
\end{split}
\end{equation}

We will present the $H^\infty$ analysis and synthesis results in the following sections.

\section{Robust $H^\infty$ Analysis}\label{section4}
To proceed the robust $H^\infty$ analysis, we first recall the strict bounded real lemma for a general quantum system. Let us consider the following quantum system of the form (\ref{linear}):
\begin{equation}\label{2}
\begin{split}
dx(t)=&Ax(t)dt+[B\quad G]\\
      &\times[dw(t)^T\quad dv(t)^T]^T;\\
dz(t)=&Cx(t)dt+Ddv(t).\\
\end{split}
\end{equation}

From Corrollary 4.5 in \cite{matt2008}, we know that
this quantum stochastic system (\ref{2}) is strictly bounded real with disturbance attenuation $g>0$ if and only if there exists a positive definite symmetric matrix $X>0$ such that
\begin{equation}\label{3}
\begin{split}
&A^TX+XA+C^TC+g^{-2}XBB^TX<0.
\end{split}
\end{equation}

When there is uncertainty $\Delta$ in the state matrix of (\ref{2}), the system is of the form
\begin{equation}\label{4}
\begin{split}
dx(t)=&(A+2\Theta E^T\Delta E)x(t)dt+[B\quad G]\\
      &\times[dw(t)^T\quad dv(t)^T]^T;\\
dz(t)=&Cx(t)dt+Ddv(t).\\
\end{split}
\end{equation}
%
In order to guarantee an $H^\infty$ performance of (\ref{4}) for all admissible parameter uncertainties, we incorporate $\Delta$ in (\ref{3}).

\begin{definition}
The quantum stochastic system (\ref{4}) is robustly strict bounded real with disturbance attenuation $g>0$ if there exists a positive definite symmetric matrix $X>0$ such that
\begin{equation}\label{5}
\begin{split}
&(A+2\Theta E^T\Delta E)^TX+X(A+2\Theta E^T\Delta E)+C^TC\\
&+g^{-2}XBB^TX<0
\end{split}
\end{equation}
\end{definition}
for all the admissible $\Delta$.

We are now in the position to introduce a scaled system for establishing a connection between the coherent robust $H^\infty$ control problem and a scaled $H^\infty$ control problem. The scaled system without parameter uncertainty has the following form:
\begin{equation}\label{6}
\begin{split}
dx(t)=&Ax(t)dt+[[2\sqrt{\epsilon}\Theta E^T\quad g^{-1}B]\quad G]\\
      &\times[d\overline{w}(t)^T\quad dv(t)^T]^T;\\
d\overline{z}(t)=&[\frac{1}{\sqrt{\epsilon}}E^T\quad C^T]^Tx(t)dt+[0\quad D^T]^Tdv(t),\\
\end{split}
\end{equation}
where $\epsilon>0$ is a scaling parameter.

Then we show the connection between the robust strictly bounded realness of the system (\ref{4}) and the strictly bounded realness of the system (\ref{6}) in the following lemma.
\begin{lemma}\label{connection} (see Lemma 3.1 of \cite{lihua1992})
Let the constant $g>0$ be given. Then there exists a matrix $X>0$ such that
\begin{equation}
\begin{split}
&(A+2\Theta E^T\Delta E)^TX+X(A+2\Theta E^T\Delta E)+C^TC\\
&+g^{-2}XBB^TX<0
\end{split}
\end{equation}
for all $\Delta$ satisfying $\Delta=\Delta^T$ and $\Delta^2\leq I$ if there exists a constant $\epsilon>0$ such that
\begin{equation}
\begin{split}
&A^TX+XA+g^{-2}XBB^TX+4\epsilon X\Theta E^TE\Theta^T X\\
&+\frac{1}{\epsilon}E^TE+C^TC<0.
\end{split}
\end{equation}
\end{lemma}

According to Definition 1, Lemma 1 and (\ref{3}), we have the following corollary.
\begin{corollary}
The system (\ref{4}) is robustly strict bounded real with disturbance attenuation $g>0$ if there exists a scaling parameter $\epsilon>0$ such that the system (\ref{6}) is strictly bounded real with unitary disturbance attenuation.
\end{corollary}
\section{Robust $H^\infty$ Controller Synthesis}\label{section5}
In this section, we present a procedure to design a coherent robust $H^\infty$ controller for an uncertain linear quantum system.
\subsection{The Closed-Loop System}
We consider a plant described by noncommutative stochastic models in the following form
\begin{equation}\label{1.1}
\begin{split}
dx(t)=&(A+2\Theta E^T\Delta E)x(t)dt+[B_0\quad B_1\quad B_2]\\
      &\times[dv(t)^T\quad dw(t)^T\quad du(t)^T]^T; x(0)=x_0\\
dz(t)=&C_1x(t)dt+D_{12}du(t)\\
dy(t)=&C_2x(t)dt+[D_{20}\quad D_{21}\quad 0]\\
      &\times[dv(t)^T\quad dw(t)^T\quad du(t)^T]^T.
\end{split}
\end{equation}
Here, $w(t)$ represents a disturbance signal and $v(t)$ represents any additional quantum noise. The signal $u(t)$ is a control input of the form $du(t)=\beta_u(t)dt+d\tilde{u}(t)$ where $\tilde{u}(t)$ is the noise part of $u(t)$ and $\beta_u(t)$ is an adapted process.

The corresponding scaled $H^\infty$ control system is of the form
\begin{equation}\label{1.6}
\begin{split}
dx(t)=&Ax(t)dt+[B_0\quad [2\sqrt{\epsilon}\Theta E^T\quad g^{-1}B_1]\quad B_2]\\
      &\times[dv(t)^T\quad d\overline{w}(t)^T\quad du(t)^T]^T; x(0)=x_0\\
d\overline{z}(t)=&[\frac{1}{\sqrt{\epsilon}}E^T\quad C_1^T]^Tx(t)dt+[0\quad D_{12}^T]^Tdu(t)\\
dy(t)=&C_2x(t)dt+[D_{20}\quad [0\quad g^{-1}D_{21}]\quad 0]\\
      &\times[dv(t)^T\quad d\overline{w}(t)^T\quad du(t)^T]^T.
\end{split}
\end{equation}

Controllers are assumed to be noncommutative stochastic systems of the form
\begin{equation}\label{controller}
\begin{split}
d\xi(t)=&A_K\xi(t)dt+[B_{K1}\quad B_K]\\
      &\times[dv_K(t)^T\quad dy(t)^T]^T; \xi(0)=\xi_0\\
du(t)=&C_K\xi(t)dt+[B_{K0}\quad 0]\\
      &\times[dv_K(t)^T\quad dy(t)^T]^T
\end{split}
\end{equation}
where $\xi(t)=[\xi_1(t)\quad ... \quad \xi_{n_K}(t)]^T$ is a vector of self-adjoint controller variables.

\begin{theorem}
Let $g>0$ be a prescribed level of disturbance attenuation and a given linear dynamic controller is described in (\ref{controller}). Then the system (\ref{1.1}) is robustly strict bounded real with disturbance attenuation $g>0$ via the output feedback controller (\ref{controller}) if there exists a constant $\epsilon>0$ such that the closed loop system corresponding to (\ref{1.6}) and (\ref{controller}) is strictly bounded real with unitary disturbance attenuation.
\end{theorem}
\emph{Proof}: By interconnecting (\ref{1.1}) and (\ref{controller}), and making the identification $\beta_u(t)=C_K\xi(t)$, the closed loop system is of the form
\begin{equation}\label{closed_loop}
\begin{split}
d\eta(t)=&\left[\begin{array}{c c} A+2\Theta E^T\Delta E & B_2 C_K\\B_K C_2&A_K\end{array}\right]\eta(t)dt\\
&+\left[\begin{array}{c c} B_0 & B_2 B_{K0}\\B_K D_{20}&B_{K1}\end{array}\right]\left[\begin{array}{c} dv(t)\\dv_K(t)\end{array}\right]\\
&+\left[\begin{array}{c} B_1\\B_K D_{21}\end{array}\right]dw(t)\\
dz(t)=&\left[\begin{array}{c c} C_1&D_{12}C_K\end{array}\right]\eta(t)dt\\
&+\left[\begin{array}{c c} 0&D_{12}B_{K0}\end{array}\right]\left[\begin{array}{c} dv(t)\\dv_K(t)\end{array}\right]\\
\end{split}
\end{equation}
where $\eta(t)=\left[\begin{array}{c c} x(t)^T&\xi(t)^T\end{array}\right]^T$.

We can also write it as
\begin{equation}\label{3.1}
\begin{split}
d\eta(t)=&(\tilde{A}+2\tilde{\Theta}\tilde{E}^T\Delta\tilde{E})\eta(t)dt+\tilde{B}dw(t)+\tilde{G}d\zeta(t)\\
dz(t)=&\tilde{C}\eta(t)dt+\tilde{H}d\zeta(t)\\
\end{split}
\end{equation}
where
\begin{equation}\label{2.1}
\begin{split}
\zeta(t)&=\left[\begin{array}{c} v(t)\\v_K(t)\end{array}\right]; \quad \tilde{A}=\left[\begin{array}{c c} A& B_2 C_K\\B_K C_2&A_K\end{array}\right]\nonumber\\
\tilde{\Theta}&=\left[\begin{array}{cc} \Theta &0\\0&0\end{array}\right];\quad \tilde{E}=\left[\begin{array}{c c} E&0\end{array}\right]\\
\tilde{B}&=\left[\begin{array}{c} B_1\\B_K D_{21}\end{array}\right];\quad
 \tilde{G}=\left[\begin{array}{c c} B_0 & B_2 B_{K0}\\B_K D_{20}&B_{K1}\end{array}\right]\nonumber\\
 \tilde{C}&=\left[\begin{array}{c c} C_1&D_{12}C_K\end{array}\right];\quad \tilde{H}=\left[\begin{array}{c c} 0&D_{12}B_{K0}\end{array}\right].
\end{split}
\end{equation}
Also, the closed loop system of (\ref{1.6}) with controller (\ref{controller}) is of the form
\begin{equation}\label{scaled1.6}
\begin{split}
d\eta(t)=&\tilde{A}\eta(t)dt+\left[\begin{array}{c c} 2\sqrt{\epsilon}\tilde{\Theta}\tilde{E}^T&g^{-1}\tilde{B}\end{array}\right]d\overline{w}(t)+\tilde{G}d\zeta(t)\\
d\overline{z}(t)=&\left[\begin{array}{c} \frac{1}{\sqrt{\epsilon}}\tilde{E}\\\tilde{C}\end{array}\right]\eta(t)dt+\left[\begin{array}{c} 0\\\tilde{H}\end{array}\right]d\zeta(t)\\
\end{split}
\end{equation}
where $\tilde{A},\tilde{B}, \tilde{\Theta},\tilde{E},\tilde{C},\tilde{G},\tilde{H}$ are the same as in (\ref{3.1}).
Therefore, the desired result follows immediately from Corollary 1.
\hfill $\Box$
\begin{remark}
Essentially, in order to solve the $H^\infty$ controller synthesis problem for the uncertain quantum system (\ref{1.1}), we need to solve the scaled $H^\infty$ problem (\ref{1.6}) via an existing $H^\infty$ control technique.
\end{remark}
\subsection{$H^\infty$ Control Objective}
For a given disturbance attenuation parameter $g>0$, the $H^\infty$ control objective is to find a quantum controller of the form (\ref{controller}) for the uncertain quantum system (\ref{1.1}) such that the closed-loop system satisfies
\begin{equation}\label{hcontrol}
\begin{split}
&\int_0^t\langle \beta_z(s)^T\beta_z(s)+\tilde{\epsilon}\eta(s)^T\eta(s)\rangle ds\\
&\leq(g^2-\tilde{\epsilon})\int_0^T\langle \beta_w(s)^T\beta_w(s)\rangle ds+\mu_1+\mu_2t,\forall t>0
\end{split}
\end{equation}
for some real constants $\tilde{\epsilon},\mu_1,\mu_2>0$. Therefore, the controller bounds the effect of the `energy' in the disturbance signal $\beta_w(t)$ on the `energy' of the error signal $z(t)$. We should also notice that if the closed-loop system (\ref{closed_loop}) is robustly strict bounded real with disturbance attenuation $g$, it then satisfies the $H^\infty$ control synthesis objective (\ref{hcontrol}).
\subsection{Necessary and Sufficient Conditions}
In this subsection, we aim to design a coherent controller for the scaled system (\ref{1.6}) to satisfy $H^\infty$ performance for unitary attenuation. Hence, based on Theorem 1, the desired coherent controller can also guarantee the uncertain quantum system (\ref{1.1}) achieve $H^\infty$ objective for a given disturbance attenuation $g$. The necessary and sufficient conditions for the existence of a specific type of controller are given. Moreover, the explicit formulas for $A_K$, $B_K$ and $C_K$ are also presented.

Now we present our controller design method. Firstly, the scaled system (\ref{1.6}) is required to satisfy the following assumptions.

\begin{assumption}\label{assumption1}
 $\quad$\\
1) $D_{12}^TD_{12}=E_1>0.$\\
2) $g^{-2}D_{21}D_{21}^T=E_2>0.$\\
3) The matrix $\left[\begin{array}{c c} A-iwI&B_2\\\frac{1}{\sqrt{\epsilon}}E&0\\C_1&D_{12}\end{array}\right]$ is full column rank for all $w\geq0$.\\
4)The matrix $\left[\begin{array}{c c c} A-iwI&2\sqrt{\epsilon}\theta E^T&g^{-1}B_1\\C_2&0&g^{-1}D_{21}\end{array}\right]$ is full row rank for all $w\geq0$.\\
\end{assumption}

The solution to the $H^\infty$ control problem for the scaled system (\ref{1.6}) is given in terms of the following pair of algebraic Riccati equations:
\begin{equation}\label{riccati1}
\begin{split}
 &(A-B_2E_1^{-1}D_{12}^TC_1)^TX+X(A-B_2E_1^{-1}D_{12}^TC_1)\\
 &+X(4\epsilon\theta E^TE\theta^T+g^{-2}B_1B_1^T-B_2E_1^{-1}B_2^T)X\\
 &+\frac{1}{\epsilon}E^TE+C_1^TC_1-C_1^TD_{12}E_1^{-1}D_{12}^TC_1=0,
\end{split}
\end{equation}
\begin{equation}\label{riccati2}
\begin{split}
&(A-g^{-2}B_1D_{21}^TE_2^{-1}C_2)Y+Y(A-g^{-2}B_1D_{21}^TE_2^{-1}C_2)^T\\
&+Y(\frac{1}{\epsilon}E^TE+C_1^TC_1-C_2^TE_2^{-1}C_2)Y\\
&+4\epsilon\theta E^TE\theta^T+g^{-2}B_1B_1^T-g^{-4}B_1D_{21}^TE_2^{-1}D_{21}B_1^T=0,
\end{split}
\end{equation}
where $X$ and $Y$ are positive-definite symmetric matrices. The solutions to these Riccati equations need to satisfy the following assumption.
\begin{assumption}\label{assumption2}
$\quad$\\
1) $A-B_2E_1^{-1}D_{12}^TC_1+(4\epsilon\theta E^TE\theta^T+g^{-2}B_1B_1^T-B_2E_1^{-1}B_2^T)X$
is a stability matrix.\\
2) $A-g^{-2}B_1D_{21}^TE_2^{-1}C_2+Y(\frac{1}{\epsilon}E^TE+C_1^TC_1-C_2^TE_2^{-1}C_2)$
is a stability matrix.\\
3) The matrix $XY$ has a spectral radius strictly less than one.
\end{assumption}

It will be shown that if the solution to the Riccati equations (\ref{riccati1}), (\ref{riccati2}) satisfies Assumption \ref{assumption2}, then a quantum controller of the form (\ref{controller}) will solve the coherent $H^\infty$ control problem where its system matrices are constructed from the Riccati solutions as below:

\begin{equation}\label{controller_solution}
\begin{split}
A_K=&A+B_2C_K-B_KC_2+4\epsilon\theta E^TE\theta^T X\\
    &+g^{-2}B_1B_1^TX-g^{-2}B_KD_{21}B_1^TX\\
B_K=&(I-YX)^{-1}(YC_2^T+g^{-2}B_1D_{21}^T)E_2^{-1}\\
C_K=&-E_1^{-1}(B_2^TX+D_{12}^TC_1).
\end{split}
\end{equation}

Now, we present our main result on coherent robust $H^\infty$ controller synthesis.
\begin{theorem}
\emph{Necessity.} Consider the system (\ref{1.6}) and suppose that Assumption \ref{assumption1} is satisfied. If there exists a controller of the form (\ref{controller}) such that the resulting closed-loop system (\ref{scaled1.6}) is strictly bounded real with unitary disturbance attenuation, then the Riccati equations (\ref{riccati1}) and (\ref{riccati2}) will have stabilizing solutions $X\geq0$ and $Y\geq0$ satisfying Assumption \ref{assumption2}.

\emph{Sufficiency.} Suppose the Riccati equations (\ref{riccati1}) and (\ref{riccati2}) have stabilizing solutions $X\geq0$ and $Y\geq0$ satisfying Assumption \ref{assumption2}. If the controller (\ref{controller}) is such that the matrices $A_K,B_K,C_K$ are as defined in (\ref{controller_solution}), then the resulting closed-loop system (\ref{scaled1.6}) is strictly bounded real with unitary disturbance attenuation.
\end{theorem}
\subsection{Physical Realization of Controllers}
As can be seen, an $H^\infty$ controller defined by the matrices $A_K,B_K,C_K$ in (\ref{controller_solution}) is not always physically realizable, that is, $A_K,B_K,C_K$ may not satisfy the relationships (\ref{preserve}) and (\ref{preserve2}). In order to guarantee the physical realizability condition, we need the following theorem.
\begin{theorem}\label{physical}(See Theorem 5.5 of \cite{matt2008})\\
Assume $F_y=D_{20}F_vD_{20}^T+D_{21}F_wD_{21}^T$ is canonical. Let ${A_K,B_K,C_K}$ be an arbitrary triple (such as given by (\ref{controller_solution})), and the controller commutation matrix is canonical $\Theta_K$. Then there exists controller parameters $B_{K0}$, $B_{K1}$, and the controller noise $v_K$ such that the controller (\ref{controller}) is physically realizable. In particular, $2i\Theta_K=(\xi(t)\xi(t)^T-(\xi(t)\xi(t)^T))^T$ for all $t\geq0$ whenever $2i\Theta_K=(\xi(0)\xi(0)^T-(\xi(0)\xi(0)^T))^T$.
\end{theorem}

To conclude, Theorem \ref{physical} shows that it is always possible to find a physically realizable controller given the matrices $A_K,B_K,C_K$.
\section{ILLUSTRATIVE EXAMPLE}\label{illustrative example}
In this section, we consider an example which illustrates the use of Theorem 1 and Theorem 2. We consider an extension of one of the examples used in \cite{matt2008} and \cite{aline2010}. This model has also been demonstrated by an experiment \cite{experiment}. As shown in Figure 1, an optical cavity is resonantly coupled to three optical channels $v,w,u$. The control objective is to attenuate the effect of the disturbance $w$ on the output $z$. The nominal quantum system is described by the evolution of its annihilation operator $a$ (representing a standing wave).
\begin{figure}[htb]
       \centering
        \includegraphics[width=0.35  \textwidth]{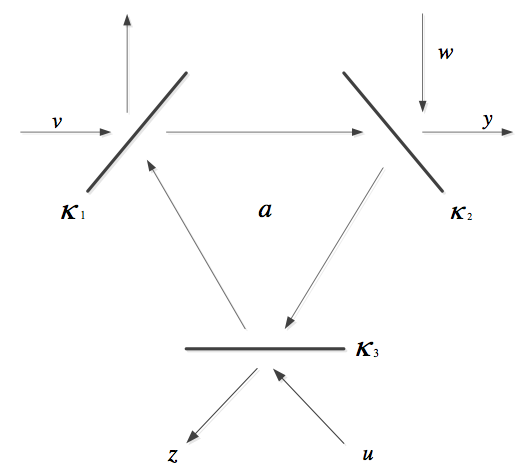}
        \caption{Optical cavity system}
        \label{fig1}
\end{figure}
In this example, we consider a detuned cavity. The uncertainty $\Delta$ represents the ``detuning" and describes the difference between the nominal external field frequency and the cavity mode frequency. The dynamics of the system are in the following form
\begin{equation}\label{annihilation}
\begin{split}
da(t)=&(-\frac{\gamma}{2}-2i\Delta)a(t)dt-\sqrt{\kappa_1}dA_1(t)-\sqrt{\kappa_2}dA_2(t)\\
 &-\sqrt{\kappa_3}dA_3(t);\\
da^*(t)=&(-\frac{\gamma}{2}+2i\Delta)a^*(t)dt-\sqrt{\kappa_1}dA_1^*(t)-\sqrt{\kappa_2}dA_2^*(t)\\
&-\sqrt{\kappa_3}dA_3^*(t);\\
dB_2(t)=&\sqrt{\kappa_2}a(t)dt+dA_2(t);\\
dB_3(t)=&\sqrt{\kappa_3}a(t)dt+dA_3(t).
\end{split}
\end{equation}
Here $\gamma=\kappa_1+\kappa_2+\kappa_3$. Also, $A_1(t),A_2(t),A_3(t)$ represent the input fields in channels $v,w,u$ respectively and $B_2(t),B_3(t)$ stand for the output fields in channels $w,u$ respectively.
The system (\ref{annihilation}) can be written in real quadrature form (\ref{1.1}) with the following system matrices:
\begin{equation}
\begin{split}
A=&-\frac{\gamma}{2}I+2J\Delta;B_0=-\sqrt{\kappa_1}I\nonumber\\
B_1&=-\sqrt{\kappa_2}I;B_2=-\sqrt{\kappa_3}I\nonumber\\
C_1&=\sqrt{\kappa_3}I;D_{12}=I\nonumber\\
C_2&=\sqrt{\kappa_2}I;D_{21}=I.
\end{split}
\end{equation}
Here, $x_1(t)=q(t)=a(t)+a^\ast(t)$ and $x_2(t)=p(t)=(a(t)-a^\ast(t))/i$. The commutation relation for this plant is described by $\Theta_p=J$ and the quantum noises $v,w$ have Hermitian Ito matrices $F_v=F_w=I+iJ$.

We choose the total cavity decay rate $\gamma=12$ and the coupling coefficients $\kappa_1=6.5, \kappa_2=5,\kappa_3=0.5$. The required disturbance attenuation constant $g=0.35$ and uncertainty range is $-1\leq\Delta\leq1$. By applying Theorem 2 to the uncertain quantum system, we get the required solutions of Riccati equations (\ref{riccati1}), (\ref{riccati2}) that satisfy Assumption 2 with $X=0.0038I$, $Y=14.0783I$. The corresponding controller matrices are given by
\begin{equation}
A_K=-34.9604I, B_K=13.5894I,C_K=-0.7058I.\nonumber
\end{equation}
Since the controller system is required to be physical realizable, we have the following form
\begin{equation}
\begin{split}
&B_{K1}=\left[\begin{array}{c c c c} 0.7058&0&8&-8\\0&0.7058&-8&-6.4062\end{array}\right],\\
 &B_{K0}=[I\quad 0].\nonumber
\end{split}
\end{equation}

To make a performance comparison between the method in this paper and the method proposed in \cite{matt2008}, we apply the approach in \cite{matt2008} and get the following results:
\begin{equation}
\begin{split}
&X=Y=0_{2\times2},\\
& A_K=-0.5I, B_K=-2.2361I, C_K=-0.7071I\\
&B_{K1}=\left[\begin{array}{c c c c} 0.7071&0&-1&1\\0&0.7071&1&3.5\end{array}\right],\\
 &B_{K0}=[I\quad 0].\nonumber
\end{split}
\end{equation}

\begin{figure}[htb]
       \centering
        \includegraphics[width=0.45  \textwidth]{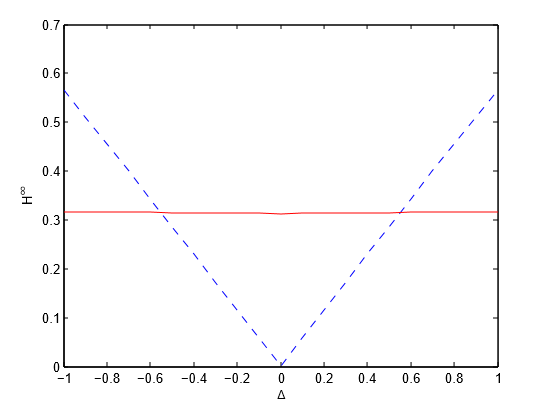}
        \caption{$H^\infty$ norm of the closed loop systems}
        \label{fig1}
\end{figure}
For the same uncertain quantum system as given before, we can make a performance comparison between the method in \cite{matt2008} where uncertainty is not considered in the controller design and the coherent robust $H^\infty$ controller presented in this paper. Figure 2 shows how the $H^\infty$ norm of the closed-loop system changes as the uncertainty varies. The dotted line shows the performance of the closed loop system with the coherent controller used in \cite{matt2008}, while the solid line describes the performance with a coherent robust controller presented in this paper. As can be seen from Figure 2, the controller in \cite{matt2008} performs better when the uncertainty variation is small, while as the uncertainty increases, the controller in \cite{matt2008} is worse than the robust controller in this paper. In the meanwhile, as uncertainty varies, the $H^\infty$ norm with the method in this paper does not change much and leads to a closed loop system having improved performance.

\section{CONCLUSION}\label{conclusion}
In this paper, we have considered a class of uncertain linear quantum systems subject to quadratic perturbations in the system Hamiltonian. For this class of given quantum systems, we have built a relationship between a coherent robust $H^\infty$ control problem and a scaled $H^\infty$ control problem without parameter uncertainty. Then, we used Riccati equations to formulate a linear dynamic output feedback quantum controller to the given quantum system to make the system robustly stable and also satisfy a prescribed level of disturbance attenuation. We also provided an optical cavity example to demonstrate the method we presented in this paper and showed that our method in this paper has improved performance over the previous result in \cite{matt2008} without considering uncertainty. In the future, we could include unknown perturbations in the system coupling operator for the uncertain quantum systems and develop a systematic $H^\infty$ control approach to deal with this kind of quantum models.


\end{document}